\documentclass[%
preprint,
 amsmath,amssymb,
 aps,
]{revtex4-2}
\usepackage{fancyhdr} % Required for custom footers

\usepackage{graphicx}% Include figure files
\usepackage{dcolumn}% Align table columns on decimal point
\usepackage{bm}% bold math

\makeatletter
\def\blfootnote{\gdef\@thefnmark{}\@footnotetext}
\makeatother

\begin{document}

% \preprint{APS/123-QED}

\title{Topological constraints suppress shear localization in granular chain ensembles}

\author{\textnormal{Palash~Sarate}$^{*,1}$,~\textnormal{Mohd.~Ilyas~Bhat}$^{*,2}$,\textnormal{~Tejas~G.~Murthy}$^{2}$,~\textnormal{Prerna~Sharma}$^{1}$}

\affiliation{Department of Physics, Indian Institute of Science, Bangalore 560012, India$^{1}$}

\affiliation{Department of Civil Engineering, Indian Institute of Science, Bangalore 560012, India$^{2}$}

\blfootnote{$^{*}$Contributed equally to this work.}

% \date{\today}

\begin{abstract}
% context
Entangled granular systems exhibit mechanical rigidity and resistance to deformation, reminiscent of cohesive materials, due to their reduced degrees of freedom and contact friction. A quantitative understanding of how classical granular phenomena, such as shear localization and plastic flow, appear in such geometrically cohesive systems remains unknown. Here, we investigate this using granular chain ensembles subjected to direct shear tests. 
Our experiments reveal that chains longer than four beads exhibit pronounced shear hardening, which is nearly independent of the applied normal stress and is accompanied by the complete suppression of shear localization. The volume dilation of the long chain ensembles also does not vanish in the steady state. 
We complement this phenomenology, which is distinct from that of typical frictional granular ensembles, with DEM simulations. The simulations reveal that tensile forces are generated due to particles being locally jammed, characterized by a high non-covalent coordination number. Consequently, this leads to a deformation that shows a very diffuse region of localization and enhanced shear hardening. Overall, our study highlights that granular chains provide a systematic route to map how connectivity constraints impact flow properties and mechanical rigidity.
\end{abstract}

\maketitle

\section*{Significance statement}
The addition of constraints in one dimension to particles allows the formation of free-standing columns rather than collapse into heaps. When 1D constrained granular ensembles - i.e., granular chains are sheared, not only does the stiffness significantly increase, but localization into narrow shear zones is suppressed. Through a battery of experiments and simulations, we investigate the kinematics of deformation in chain ensembles and find that the development of tensile forces plays a key role.  Our results are especially relevant in granular metamaterials, earthquake-resistant geostructures, and 3D printing. 
\section{Introduction}
The control of the rheological properties of granular media plays a key role in the achievement of efficient industrial processes and the prevention of geological disasters such as avalanches and landslides. 
It is well established that the mechanical response of an ensemble of dry frictional granular spheres exhibits a linear elastic behavior only up to very small strains, post which the material deforms plastically, and often the dense ensemble shows a strain softening response ~\cite{Roscoe1958,zhou2017dem,Rao2008} to reach a critical state (or the terminal state). Such deformations are rarely homogeneous, and often regions of localized deformations emerge. Specifically, such bifurcation like instabilities cause deformation to become non-uniform within the system even though the whole system is sheared under a homogeneous stress state~\cite{Kothari2017,Jagla2024,Mandal2021}. 
That is, regions of deformation and flow become localized within a relatively thin zone within the material, which are often referred to as ``shear bands". 
Shear bands are mesoscale features that represent the breakdown of homogeneity, a key feature required to associate bulk properties of viscosity and elasticity with a system~\cite{Berthier2025,Bardet1992,Zhu2017,Cao2018}. They often consist of buckled force chains or rotating particle clusters and serve as the primary failure mechanism in dense granular materials~\cite{Tordesillas2007,Tordesillas2010,Guo2012,Juanes2024, Karuriya2023}. 
 
For a typical granular material, the orientation of these shear bands can be estimated with the Mohr-Coulomb failure criterion, which depends on the angle of internal friction and cohesion in the system~\cite{Andreotti2013}. 
Traditionally, cohesion is understood to arise from adhesive forces between particles. For example, stored powders, natural soil deposits gain cohesion in the presence of excess moisture and organic matter~\cite{Shakeel2022,Kemper1987, Mandal2020}. However, cohesion can also emerge from the interlocking and entanglement of the particles themselves, a phenomenon termed ``geometric cohesion"~\cite{zou2009,Murphy2016,Franklin2014,Trepanier2010,Aponte2024,Jung2025}.

\begin{figure}[h]
\centering
\includegraphics[width=\linewidth]{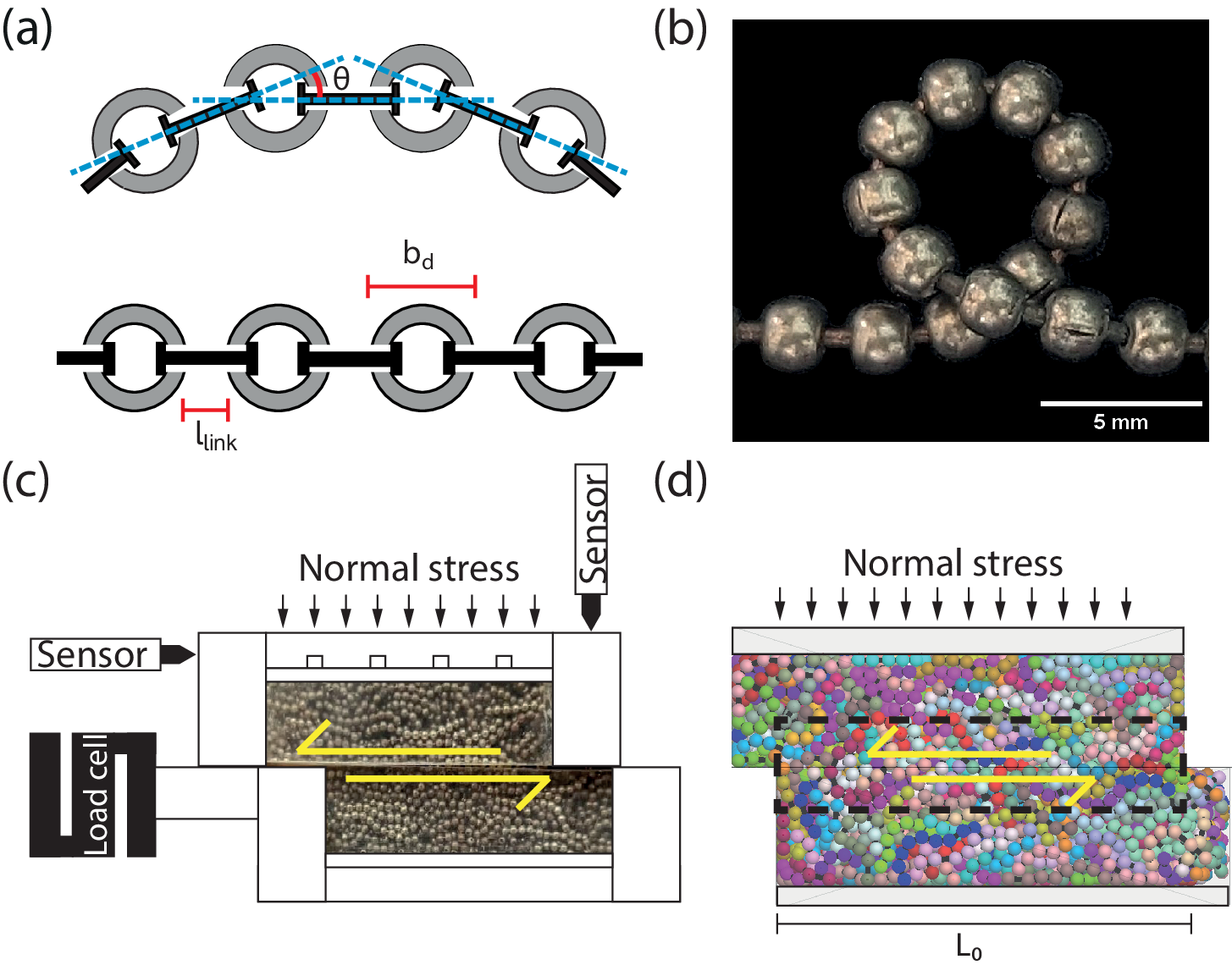}
\caption{a) Schematic of the chain geometry. b) The smallest possible loop of a granular chain comprised of 9 beads, each bead of diameter $b_d = 2~mm$, highlighting the flexibility and spacing constraints (scale bar, $5~mm$). c) Schematic of the experimental setup for direct shear testing. 
d) Simulation setup at the final state of the direct shear test simulation. The "deformation region" is indicated by the black-dashed rectangle and is defined as a zone of thickness equal to six bead diameters. $L_0$ denotes the total length of the deformation region.
}

\label{fig:expt_setup}
\end{figure}

Granular chains provide an ideal model system to study geometric cohesion, as the constraints between the linked beads and the resulting entanglement create an effective cohesion that can be tuned by altering the chain length ~\cite{1,zou2009,Brown2012,Dumont2018}. While prior work on entangled systems has largely focused on their enhanced bulk mechanical stability, a quantitative understanding of how these geometric constraints affect the meso-scale shear rheology and localization, phenomena that are hallmarks of granular media, has not emerged.  
This study addresses that gap by investigating how entanglement fundamentally alters the shear localization response. Through direct shear tests (DST) and Discrete Element Method (DEM) simulations, with setups illustrated in Fig.~\ref{fig:expt_setup}, we show that granular chain ensembles display a dramatically different behavior of shear hardening, contrasting with the shear-softening of ensembles of cohesionless spheres. We provide microscopic insights into this observed rheology, revealing how interlocking leads to qualitatively new deformation regimes.

\section{Results}
We record the ensemble response of the granular chains through a suite of direct shear tests. The test allows the application of a normal stress while localizing shear deformation in the horizontal plane.
% while forming a horizontal shear zone
The objective is to understand how localization manifests with changing chain lengths, for which we image this localization zone. 
We complement these experimental findings with insights into the microstructure through DEM simulations replicating the test. We examine the test results in the framework of granular plasticity~\cite{Rao2008,Wood1991}.

\subsection{Shear Hardening and Dilatancy}
\begin{figure}[h]
% \centering
\includegraphics[width=\linewidth]{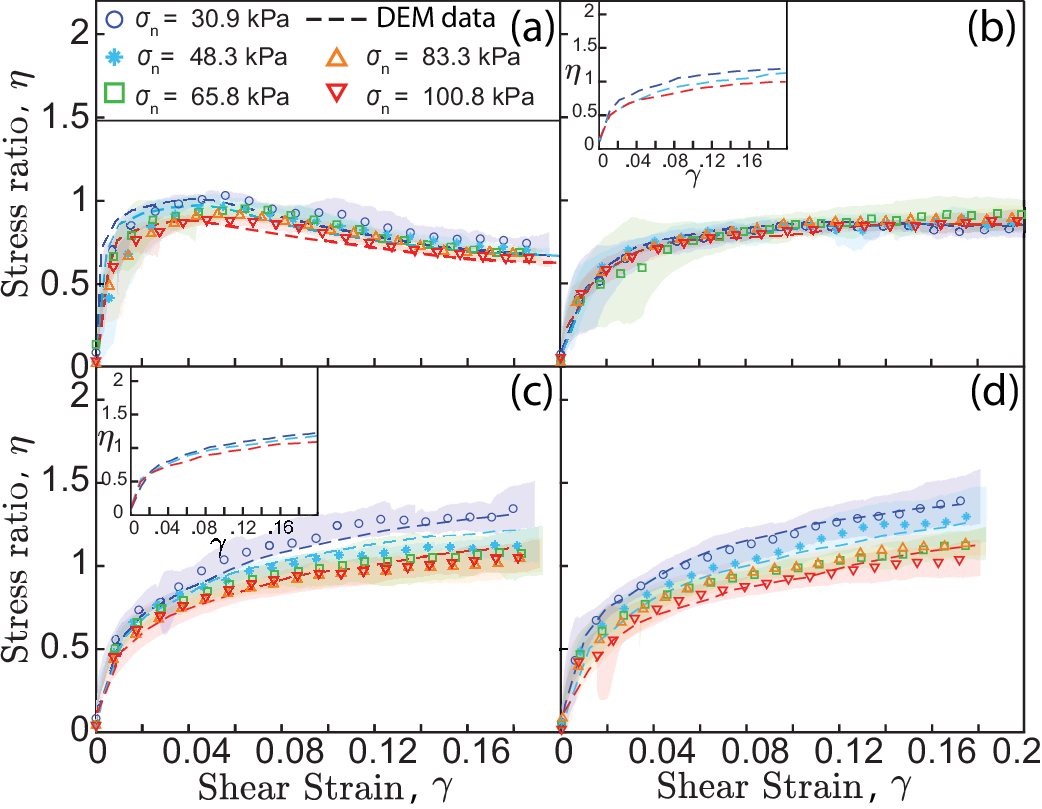}
\caption{Stress ratio-Shear strain curves from experiments and DEM simulations for different chain lengths: a) $N=1$, b) $N=4$, c) $N=12$, d) $N=24$, b.inset) $N=8$, and c.inset) $N=10$. Symbols represent experimental data, while data from DEM simulations are shown as dashed lines. The shaded region is the standard deviation in stress ratio.}
\label{fig:stressRatio_shearStrain}
\end{figure}

The stress ratio - ($\eta$) (the ratio of shear stress to normal stress), for a collection of dense granular beads ($N=1$), exhibits a peak followed by a decrease with increasing shear strain, a clear shear(strain) softening response~\cite{Roscoe1958,zhou2017dem,Rao2008}. We consistently observed this for $N=1$ across all applied normal stresses, ranging from $30.88~kPa$ to $100.75~kPa$ (Fig.~\ref{fig:stressRatio_shearStrain}a). In contrast, longer chains with ($N = 4, 12, 24, \text{ and } 48$) exhibit shear (strain) hardening, with the stress ratio increasing as the shear strain increases and ultimately saturating to a plateau only for $N=4$ (Fig.~\ref{fig:stressRatio_shearStrain}b-d, Supplementary information Fig.~S4). Representative kinematics and a direct comparison between experiment and DEM simulation for N = 1, 12 and 24 are provided in Supplementary Movies S1–S3, respectively.

% \onecolumn
% \twocolumn

\begin{figure}[h]
\includegraphics[width=\linewidth]{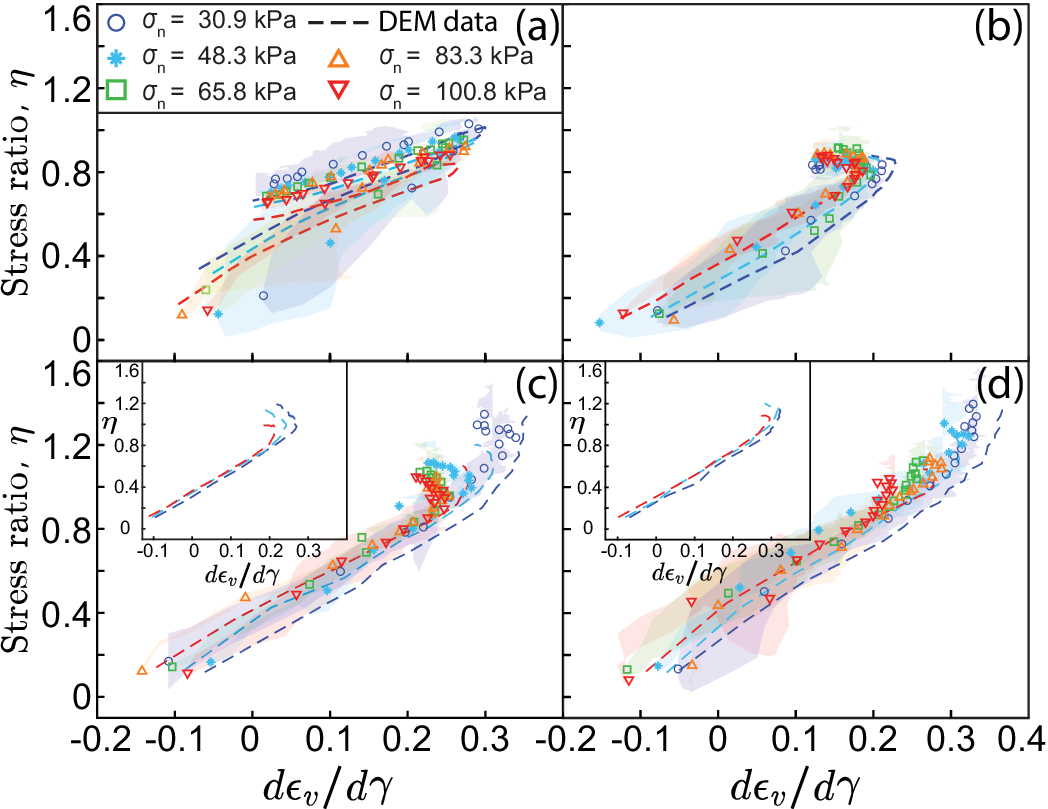}
\caption{Stress–Dilatancy relationship from experiments and DEM simulations for different chain lengths: a) $N=1$, b) $N=4$, c) $N=12$, d) $N=24$, b.inset) $N=8$, and c.inset) $N=10$, under various normal stresses, highlighting how the mechanical response changes with chain length and applied stress. Symbols represent experimental data, while data from DEM simulations are shown as dashed lines. The shaded region is the standard deviation in stress ratio.}
\label{fig:stressRatio_dilatancy}
\end{figure}

With increasing chain lengths, the stress ratio continues to increase (i.e., strain harden) with continued deformation. 
% Increasing the applied normal stress leads to a reduction in the stress ratio for all chain lengths.
An increase in  normal stress leads to a decrease in the stress ratio for all chain lengths
(Fig.~\ref{fig:stressRatio_shearStrain}). 
DEM simulations (dotted lines) accurately capture the experimental (symbols) response (Fig.~\ref{fig:stressRatio_shearStrain}) for all chain lengths used in this study. The transition from absence of strain hardening to strain hardening takes place between the chain lengths of $N = 8 \text{ and } 10$. The consequent volume change in the system during shear is measured by tracking the top of the shear box using a displacement sensor. This overall volume change is presented, normalized to the applied shear deformation, as dilatancy (Fig.~\ref{fig:stressRatio_dilatancy}).

% Experimental observations
In a typical granular ensemble, the stress ratio and dilatancy continue to increase until a peak stress state is reached, beyond which the dilation begins to decrease, eventually reaching a state of zero dilation (critical state). This behavior is clearly observed for the $N=1$ case (Fig.~\ref{fig:stressRatio_dilatancy}a) across all applied normal stresses.
In the case of the longer chains, the stress ratio and volume change continue to increase with no clear sign of a critical state (as alluded to in the previous section). The maximum dilatancy attained in the experiments increases with chain length. At higher normal stress values, dilation is suppressed for all chain lengths (Fig.~\ref{fig:stressRatio_dilatancy}b-d).

% \onecolumn
% \twocolumn

\subsection{Suppression of Shear Localization}
\begin{figure}[h]
\includegraphics[width=\linewidth]{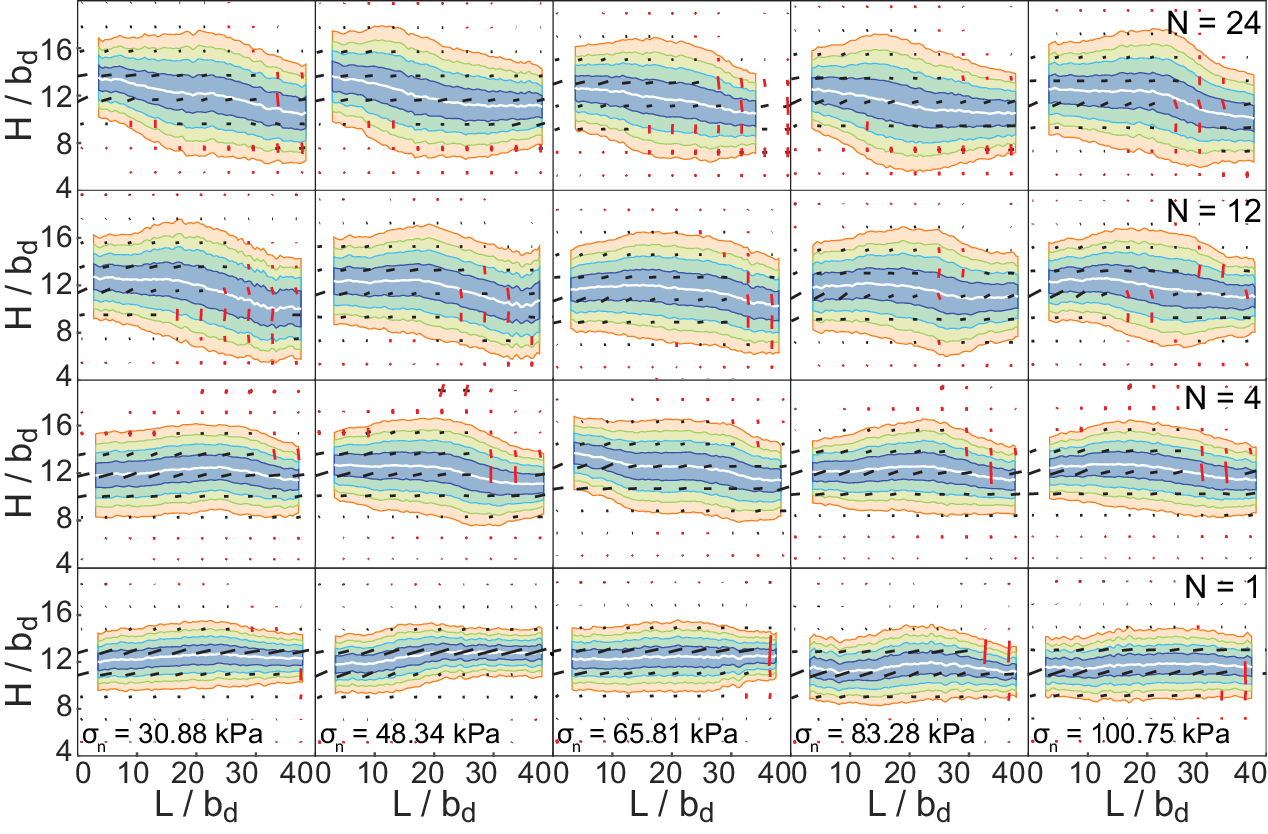}
\caption{Time-averaged representation of the shear zone for different chain lengths and applied normal stresses ($\sigma_n$) over a steady state window from $13$ to $26$ minutes. The chain length increases upwards, and $\sigma_n$ increases to the right. The color-coded regions illustrate the extent of the shear zone at different percentages of the maximum Vertical gradient of $V_x$: red - $30\%$, green - $50\%$, light blue - $70\%$, blue - $90\%$. The resulting field of dominant ZELs is overlaid over the contours of shear zone (ZEL magnitude scaled up by a factor of 6 for visibility). $L$/$b_d$ is the length along the centerline of the shear box normalized by the bead diameter.}
\label{fig:shearBand_combined}
\end{figure}

The gradient of the horizontal velocity component ($V_x$) along the height of the sample, derived from Particle Image Velocimetry (PIV) analysis, is computed to map the region of shear zones (refer Supplementary information Section.3).
We identify regions with velocity gradients within 30\% of the maximum value of the gradient as a ``shear zone". The shear zone widens as chain length increases, in effect indicating that longer chains suppress localization (Fig.~\ref{fig:shearBand_combined}).

We compute the incremental Green–Lagrange strain tensor and extract the principal strains and their orientations on the direct shear box.  Two zero extension line (ZEL) directions are derived from the eigen vectors (refer Supplementary information Section.4).  In order to identify the plane of zero normal strain (i.e. the plane of pure shear), we evaluate the ZEL direction in a local neighborhood of size \(5b_d \times 5b_d\), and overlay on the velocity gradient maps (Fig.~\ref{fig:shearBand_combined}). 

% \onecolumn
% \twocolumn

\subsection{Micromechanics}
We supplement these experiments on the direct shear box through a suite of DEM simulations to gain micromechanical insights and explain this intriguing phenomenology. The chains are modeled using two specific contact interaction mechanisms: the link contact with I-shaped links connecting the beads in the chain, and the inter-bead interaction as a frictional linear contact (refer Supplementary information Section.5). 

\begin{figure}[h]
\includegraphics[width=0.68\linewidth]{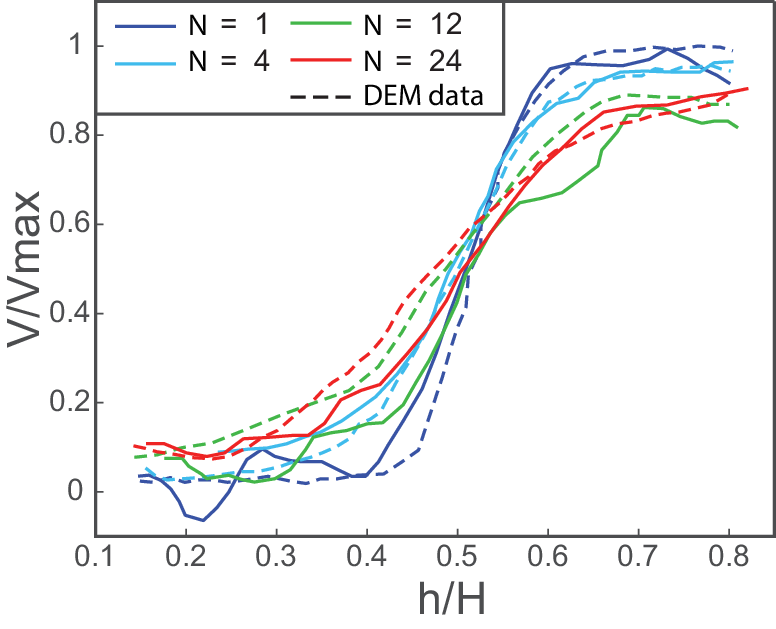}
\caption{
% (a)  DEM contact model parameters and their calibrated values. (b) 
Comparison of experimental (solid lines) and numerical (dashed lines) normalized velocity profile along the height of the sample under a normal stress of $100~kPa$. The velocities are averaged in the middle one-third width of the sample from the shear strain interval ($\gamma$) of $0.15$ to $0.16$.}
\label{fig:verif}
\end{figure}

The ensemble stress-strain response in the simulations is benchmarked against the experiments. The kinematics of the ensemble (normalized velocity profiles) from the experiments and simulations are presented in Fig.~\ref{fig:verif}). 

The gradient of the velocity profile is sigmoidal for the case of $N=1$, indicating a sharp shear zone. These velocity profiles are more diffused with increasing chain length (See Supplementary Movies S1–S3). The ensemble response emerges from a network of entangled clusters, where topological constraints prevent localized shear banding and inhibit plastic flow (otherwise typical of a purely frictional granular ensemble). These system-spanning entangled clusters are aggregates of multiple chains interlocked through knots\citep{BenNaim2001}, weaves\citep{he2025weaving}, and loops\citep{Brown2012}—whose size is governed by the persistence length (refer Supplementary information Fig.~S7). The topology of the chains constantly evolves through the shearing process, and the ensemble continues to shear harden. These clusters of chains reorient along the principal shear direction, and the inter-cluster voids continue to grow, leading to continued dilation of the ensemble. These effects are more pronounced with increased chain length as observed in our experiments.  

\begin{figure}[h]
\includegraphics[width=\linewidth]{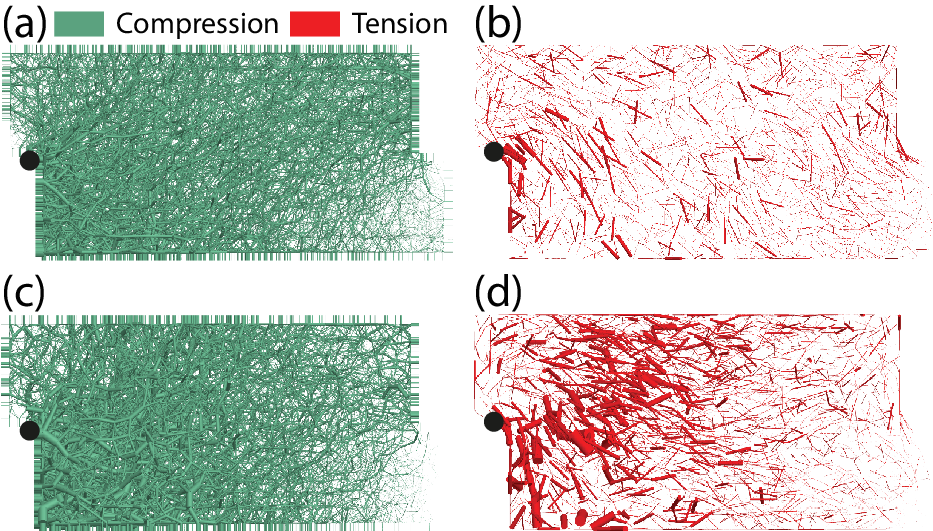}
\caption{
(a) Terminal-state compressive force chain network for a chain ensemble with \( N = 4 \).
(b) Corresponding tensile force chain network for \( N = 4 \).
(c) Terminal-state compressive force chain network for \( N = 24 \).
(d) Corresponding tensile force chain network for \( N = 24 \).
Inter-particle contact forces are visualized as cylinders connecting bead centers, with cylinder radii scaled by \( 6.3 \times 10^{-5} \) times the force magnitude. Black dots mark the ``initiation" points of the deformation region. The initial and terminal states correspond to mechanical equilibrium at the start and end of the shearing process, respectively; the terminal state is shown at a shear strain of \( \gamma = 0.2 \).
}
\label{fig:force_chain}
\end{figure}

\begin{figure}[h]
\includegraphics[width=\linewidth]{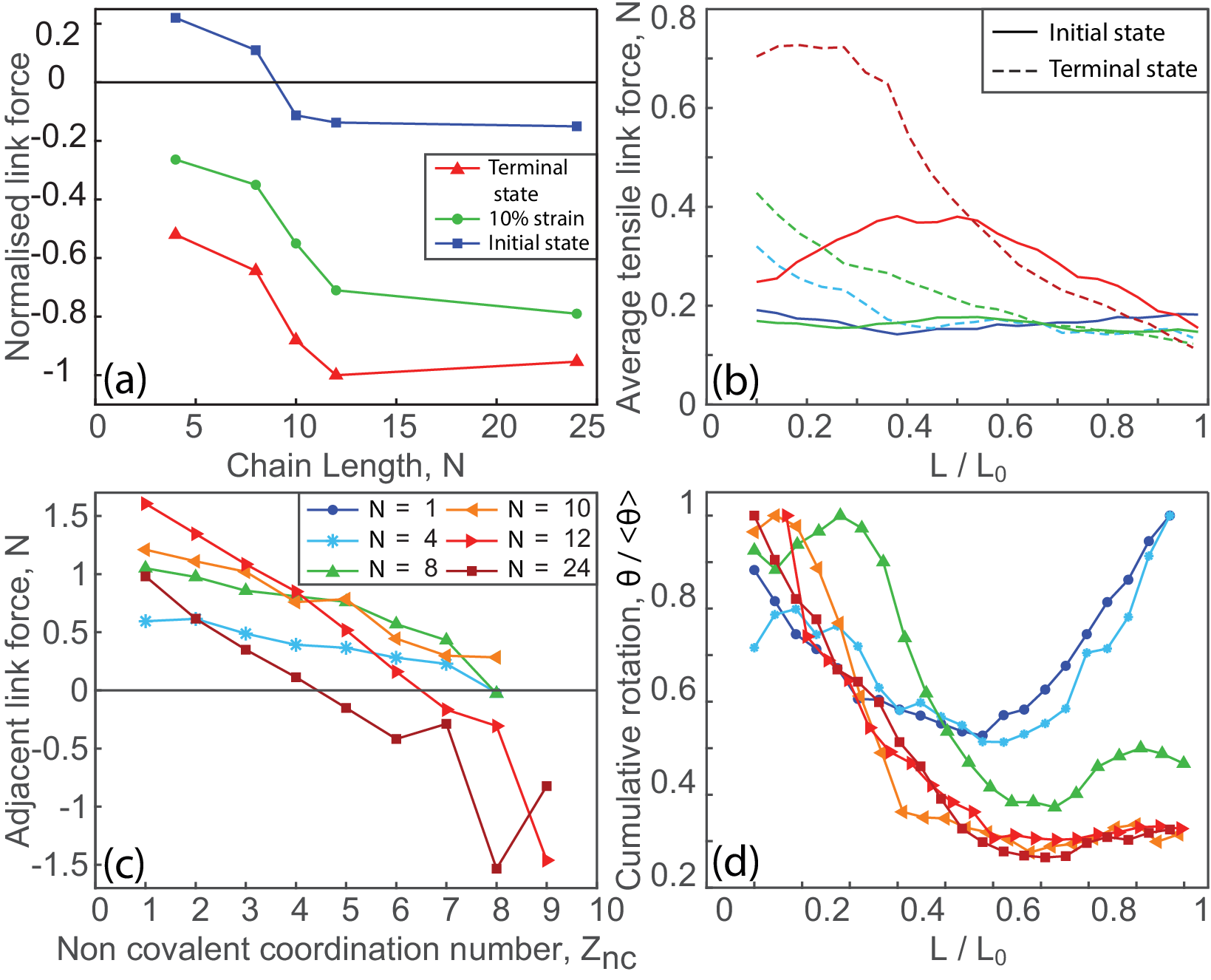}
\caption{ a) Normalized average link force in chain ensembles. Compressive forces are assigned a positive sign, while tensile forces are assigned a negative sign. The force values are normalized by the average link forces in the $N = 12$ ensemble at the terminal state. b) Absolute average tensile link force at the initial and the terminal state along the deformation region for chain ensembles of $N = 4,8, \text{ and } 24$.  c) Variation of average adjacent link force with non-covalent coordination number of the beads. d) Variation of cumulative particle rotation along the deformation region at the terminal state. The normalized distance along the deformation region, $L/L_0$, is measured along its centerline.}
\label{fig:micromechanics}
\end{figure}

Complementarily, when the force network of the chains is observed up to a terminal state, we find that with increasing shear strain, compressive forces in the ensembles continue to strengthen through frictional contacts. Tensile forces are also developed in the link contacts, especially near the initiation of the deformation region (see Fig.~\ref{fig:expt_setup}d for deformation region definition). 
We parse the compressive and tensile forces in the simulations in Fig.~\ref{fig:force_chain}. The compressive forces form a dense network spanning the entire ensemble. The tensile forces, on the other hand, radially emanate from the initiation of the deformation region (Fig.~\ref{fig:force_chain}). 
The images from the experiment reveal suppression of a characteristic shear localization with increasing chain length, and we identify the corresponding micromechanical signatures that lead to this suppression (Fig.~\ref{fig:micromechanics}b, d).
The average link tensile forces in the ensemble increase with chain length and saturate for chain lengths beyond N=12 (Fig.~\ref{fig:micromechanics}a). 
There is a jump of about $20\%$ in average normalized link tensile force from $N = 8$ ensemble to $N = 12$ ensemble (Fig.~\ref{fig:micromechanics}a).

In the case of $N = 4$ and $N = 8$, compressive link forces exist at the initial stage, and the links progressively develop tensile forces during shear (Fig.~\ref{fig:micromechanics}a). 
When we examine the links that carry only the tensile forces in the ensemble, the average tensile link forces are small (about $0.2N$ - i.e. the chains are in a slack state). Upon shearing, the chains reach a taut state and the average tensile link forces increases significantly near the initiation of the deformation region (Fig.~\ref{fig:micromechanics}b). 
The average tensile force in the links near the initiation of the deformation region is about 2 to 4 times higher than the link force at the end of the deformation region, implying heterogeneous development of tensile forces in the sample. This heterogeneity of the tensile forces increases with chain length (Fig. \ref{fig:micromechanics}b).   

The non-covalent coordination number of a bead in the chain is defined as the coordination number of the bead excluding the link contacts (only the frictional contacts). 
We define the adjacent link force as the sum of the link forces of a given bead, including both the tension-transmitting links and compressed links (Fig.~\ref{fig:micromechanics}c).
We find that the average adjacent link forces of a given bead become more tensile and increase with increasing non-covalent coordination number $Z_{nc}$(Fig.~\ref{fig:micromechanics}c). High $Z_{nc}$ implies local jamming, i.e., a particle with higher constraints and reduced degree of freedom. This observation points out a strong correlation between the tensile forces generated in the links due to the localized jamming as represented by the non-covalent coordination number of a bead (Fig.~\ref{fig:micromechanics}c). 
In conventional cohesionless granular materials, the shear zone is associated with the concentration of cumulative particle rotation and rolling, usually understood as ``frictional behavior".
For conventional cohesionless granular materials (N=1), the cumulative particle rotations are symmetric about the center of the deformation region (also for $N = 4$) (Fig.~\ref{fig:micromechanics}d).
Whereas in the case of longer chains, the cumulative particle rotation along the deformation region length is asymmetric, i.e., mostly concentrated at the start (Fig.~\ref{fig:micromechanics}d).

\begin{figure}[h]
\includegraphics[width=\linewidth]{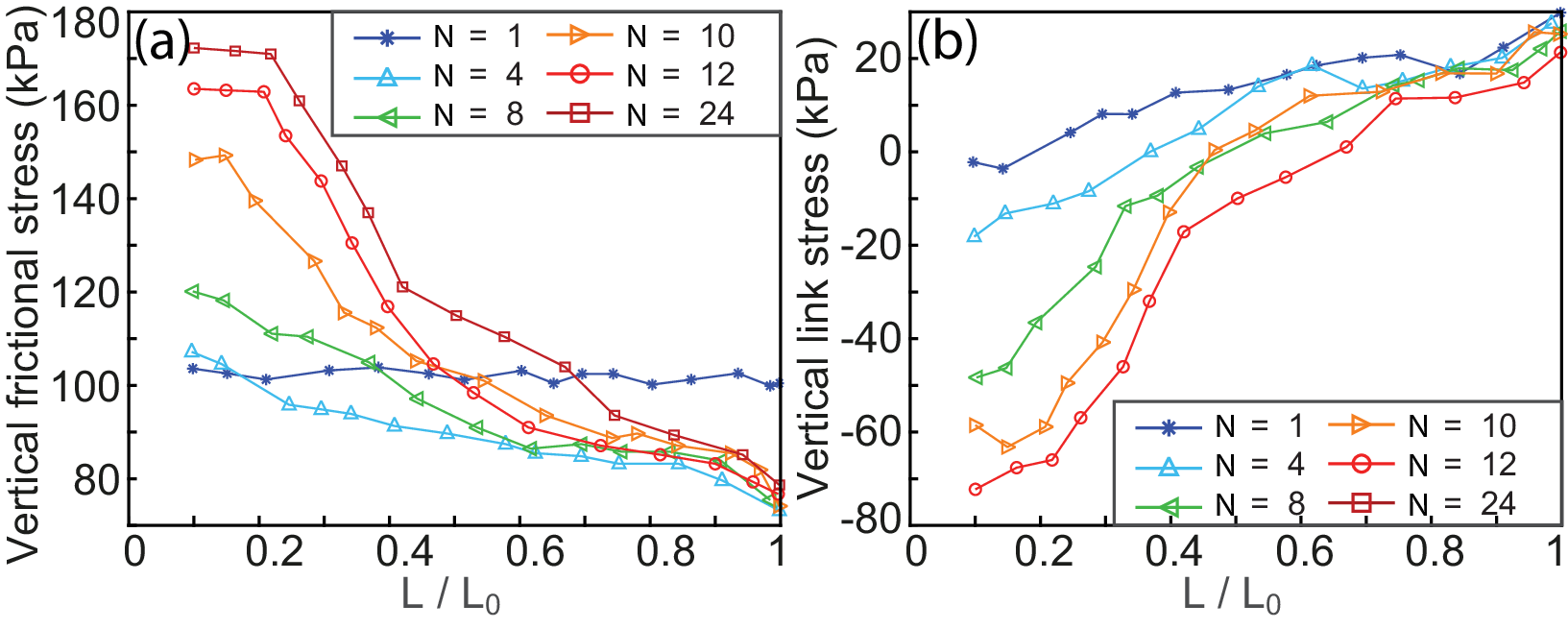}
\caption{a) Variation of compressive vertical frictional stress component and b) Variation of tensile vertical link stress component along the deformation region for ensembles of chain length $N = 1,4,8,10,12, \text{ and } 24$. All ensembles are sheared under a normal stress of $100~kPa$. The x-axis is the normalized length along the deformation region. The externally applied total vertical normal stress on the sample decomposes into a compressive frictional component and a tensile link component, with the sum of the two components conserved.}
\label{fig:stress_components}
\end{figure}

Two stress tensors are calculated by taking into account only the frictional contact force vectors or the link contact force vectors in the representative elemental volume (REV). 
REVs are selected continuously along the deformation region of the sample, and the two stress tensors are calculated for REVs along the entire length of the deformation region. The variation of vertical normal stress along the deformation region (Fig.~\ref{fig:stress_components}) is computed as the zz component (normal to the deformation region) of the stress tensor \cite{OSullivan2011}.
We present the variation of the vertical frictional stress component and the vertical link stress component along the deformation region in Figures \ref{fig:stress_components}a, b, respectively. 
The vertical frictional component decreases along the deformation region, while the corresponding vertical link stress component increases. The decomposition of the applied normal stress on the sample into a tensile link and a compressive frictional component is more pronounced in the longer-chain samples. The frictional contact stresses are still predominant when compared to the tensile link stresses in all the ensembles analyzed here.

\section*{Methods}

\section*{Experimental}
We use a direct shear test setup with box dimensions of $6 \times6~cm$ and a depth of $4~cm$ with one transparent wall. The schematic diagram of the direct shear apparatus used is shown in Fig.~\ref{fig:expt_setup}c. The lower half of the shear box lies on roller bearings and is moved horizontally at a constant velocity.
A load cell is mounted on the lower half of the box. Two displacement sensors are mounted as shown in Fig.~\ref{fig:expt_setup}c to measure the horizontal displacement of the lower half and the vertical displacement of the upper half of the shear box. Concomitantly, high-speed, high-resolution images are captured from the transparent side of the shear box. 

Granular chains made of nickel-coated brass beads (diameter $b_d = 2~mm$) connected by metallic `I-shaped' links are used. The links allow the adjacent beads to be separated by up to $1~mm$ and to have a bending angle of $\approx 40^\circ$. Chain flexibility is characterized by the minimum loop diameter of $\approx 8~mm$. We calculate the persistence length to be $\approx 12~b_d$ \cite{1}. Packing characteristics as a function of chain length are provided in Supplementary information Fig.~S1.

The shear box is set up such that the lower and upper halves of the box are perfectly aligned and separated by a uniform gap of $1~mm$. Chains are deposited into the box at its maximum packing fraction (refer Supplementary information Fig.~S1). The Shear rate is kept constant at $0.25~mm/min$.

Particle Image Velocimetry (PIV) analysis is conducted on the images captured \cite{pivlab} during the experiment. The processed data are then visualized as quiver plots and contour maps.

\section*{Simulation}
We employ Discrete Element Method (DEM) simulations to study micromechanical signatures of the ensemble response of these chains. Details of the simulations are presented in Supplementary Information Section.5. 

Links are modeled using constraints on both their extensions and bending angles. While the frictional contacts are modeled as linear springs. The two friction coefficients, i.e., sliding friction ($\mu_{s}$) and rolling friction ($\mu_{r}$), are obtained from experiments on the angle of repose \cite{bhat2023force}. The normal stiffness $k_n$, angular stiffness $k_{\theta}$, and the sample generation protocol are calibrated to reproduce the experimental stress-strain and volumetric behavior of $N = 1,4,12, \text{ and } 24$ chains. 

The simulations consist of two halves of the shear box, which are separated by a distance of half the average bead radius (as shown in Fig.~\ref{fig:expt_setup}d. The chains are deposited under gravity and tapped to achieve the required density. Normal forces are applied to the top half, and a constant horizontal displacement is provided to the bottom half (See Supplementary Movies S1–S3). The net shear stress is calculated from the unbalanced horizontal force on two halves of the shear box, and the volumetric strain is calculated from the displacement of the top wall (Fig.~\ref{fig:expt_setup}c).

\section{Discussion}
In summary, we find that granular chain ensembles display unique phenomena of shear hardening, suppression of shear localization, and continued dilation with shear deformation. These are in stark contrast with frictional granular ensembles, wherein deformation occurs over a narrow zone; the shear stresses typically show a softening response, and dilation vanishes at the steady state \citep{Roscoe1958,bolton1986strength}. 
The shear zone, where deformations are localized, becomes more diffuse with increasing chain length. This dependence on chain length manifests itself in all aspects of the bulk mechanical response. We complemented our experimental results with simulations to gain a deeper understanding of these phenomena at the microscopic level.

Simulations showed that the average link forces increase with chain length until $N = 10$ (the smallest stable loop size of the chains) and then saturate beyond that.  These tensile link forces develop preferentially near the initiation of the deformation region. The regions of high tensile link forces are also the regions of increased non-covalent coordination number, i.e., localized jamming occurs at the beginning of these diffuse shear zones (refer Supplementary information Fig.~S8). These microscopic mechanisms provide a qualitative understanding of how shear hardening appears in entangled granular chain ensembles. The development of significant tensile forces when the system is sheared implies that either the system spanning nature of entanglement must be overcome or the chain links must physically break for any shear localization to occur. Since neither of these can happen, we see that at the ensemble level, the stresses show hardening, the deformation does not show any localization, and typical signatures of entanglement, such as high coordination number, etc, continue to persist within the system. 

Our results place granular chains within the same class of connectivity-controlled materials that includes entangled polymer melts, fiber suspensions, and jammed filament networks—but in a limit where contacts are frictional and non-cohesive \citep{de1971reptation,doi1978dynamics}.  Increased correlation lengths and tension-bearing segments in soft matter systems are key signatures of entanglement \citep{mackintosh1995elasticity, larson1999structure}. Here, we observe the geometric analogue of these signatures emerging from bead-link topology alone, without the need for elasticity or chemical bonds. Within granular physics, the system provides a minimal model demonstrating how added degrees of connectivity suppress shear localization and result in shear hardening, diverging from classical frictional materials. Our work lays the foundation for examining families of chain-like materials where the constraints of connectivity, bending stiffness, and link compliance can be varied independently to systematically map the relationship between geometric cohesion and emergent rheology.

% \begin{acknowledgments}
% We wish to acknowledge the support of the author community in using
% REV\TeX{}, offering suggestions and encouragement, testing new versions,
% \dots.
% \end{acknowledgments}

\nocite{*}

\bibliography{main}% Produces the bibliography via BibTeX.

@article{Roscoe1958,
   author = {K. H. Roscoe and A. N. Schofield and C. P. Wroth},
   issn = {17517656},
   issue = {1},
   journal = {Geotechnique},
   title = {On the yielding of soils},
   volume = {8},
   year = {1958},
   pages = {22-53},
}

@article{zhou2017dem,
  author    = {W. Zhou and J. Liu and G. Ma and X. Chang},
  title     = {Three-dimensional DEM investigation of critical state and dilatancy behaviors of granular materials},
  journal   = {Acta Geotechnica},
  year      = {2017},
pages     = {527--540},
  volume    = {12},
}

@book{Rao2008,
   author = {K. Kesava. Rao and Prabhu R.. Nott},
   isbn = {9780521571661},
   publisher = {Cambridge University Press},
   title = {An introduction to granular flow},
   year = {2008}
}

@article{Kothari2017,
   author = {Konik R. Kothari and Ahmed E. Elbanna},
   issn = {24700053},
   issue = {2},
   journal = {Phys. Rev. E},
   month = {2},
   pages = {022901},
   pmid = {28297960},
   publisher = {American Physical Society},
   title = {Localization and instability in sheared granular materials: Role of friction and vibration},
   volume = {95},
   year = {2017}
}

@article{Jagla2024,
   author = {E. A. Jagla},
   issn = {1744-6848},
   issue = {3},
   journal = {Soft Matter},
   month = {1},
   pages = {588-598},
   pmid = {38131393},
   publisher = {The Royal Society of Chemistry},
   title = {From shear bands to earthquakes in a model granular material with contact aging},
   volume = {20},
   year = {2024}
}

@article{Mandal2021,
   author = {Sandip Mandal and Maxime Nicolas and Olivier Pouliquen},
   journal = {Phys. Rev. X},
   title = {Rheology of Cohesive Granular Media: Shear Banding, Hysteresis, and Nonlocal Effects},
  pages = {021017},
   volume = {11},
   year = {2021}
}

@article{Berthier2025,
   author = {Ludovic Berthier and Giulio Biroli and Lisa Manning and Francesco Zamponi},
   issn = {25225820},
   issue = {6},
   journal = {Nat. Rev. Phys.},
   month = {6},
   publisher = {Springer Nature},
   title = {Yielding and plasticity in amorphous solids},
    pages = {313},
   volume = {7},
   year = {2025}
}

@article{Bardet1992,
   author = {J. P. Bardet and J. Proubet},
   issn = {0003-6900},
   issue = {3S},
   journal = {Appl. Mech. Rev.},
   month = {3},
   publisher = {American Society of Mechanical Engineers Digital Collection},
   title = {The Structure of Shear Bands in Idealized Granular Materials},
   volume = {45},
    number = {3S},
    pages = {S118-S122},
   year = {1992}
}

@article{Zhu2017,
   author = {Huaxiang Zhu and Guillaume Veylon and François Nicot and Félix Darve},
   issn = {19648189},
   issue = {7-8},
   journal = {European Journal of Environmental and Civil Engineering},
   month = {8},
   publisher = {Taylor and Francis Ltd.},
   title = {On the mechanics of meso-scale structures in two-dimensional granular materials},
   volume = {21},
pages = {912-935},
   year = {2017}
}

@article{Tordesillas2007,
   author = {A. Tordesillas},
   issn = {14786435},
   issue = {32},
   journal = {Philosophical Magazine},
   month = {11},
   pages = {4987-5016},
   publisher = {Taylor & Francis Group},
   title = {Force chain buckling, unjamming transitions and shear banding in dense granular assemblies},
   volume = {87},
   year = {2007}
}

@article{Tordesillas2010,
   author = {Antoinette Tordesillas and David M. Walker and Qun Lin},
   issn = {15393755},
   issue = {1},
   journal = {Phys. Rev. E},
   month = {1},
   pages = {011302},
   publisher = {American Physical Society},
   title = {Force cycles and force chains},
   volume = {81},
   year = {2010}
}

@article{Guo2012,
   author = {Peijun Guo},
   issn = {18611125},
   issue = {1},
   journal = {Acta Geotechnica},
   month = {3},
   pages = {41-55},
   publisher = {Springer},
   title = {Critical length of force chains and shear band thickness in dense granular materials},
   volume = {7},
   year = {2012}
}

@book{Andreotti2013,
   author = {Bruno Andreotti and Yoël Forterre and Olivier Pouliquen},
   isbn = {9781107034792},
   month = {6},
   publisher = {Cambridge University Press},
   title = {Granular Media},
   year = {2013}
}

@article{Shakeel2022,
   author = {Ahmad Shakeel and Florian Zander and Jan Willem de Klerk and Alex Kirichek and Julia Gebert and Claire Chassagne},
   issn = {16147480},
   issue = {11},
   journal = {Journal of Soils and Sediments},
   month = {11},
   pages = {2883-2892},
   publisher = {Springer Science and Business Media Deutschland GmbH},
   title = {Effect of organic matter degradation in cohesive sediment: a detailed rheological analysis},
   volume = {22},
   year = {2022}
}

@article{Kemper1987,
   author = {W. D. Kemper and R. C. Rosenau and A. R. Dexter},
   issn = {1435-0661},
   issue = {4},
   journal = {Soil Science Society of America Journal},
   month = {7},
   pages = {860-867},
   publisher = {John Wiley \& Sons, Ltd},
   title = {Cohesion Development in Disrupted Soils as Affected by Clay and Organic Matter Content and Temperature},
   volume = {51},
   year = {1987}
}

@article{zou2009,
    author = {Ling Nan Zou  and Xiang Cheng  and Mark L. Rivers  and Heinrich M. Jaeger  and Sidney R. Nagel },
    title = {The Packing of Granular Polymer Chains},
    journal = {Science},
    volume = {326},
    number = {5951},
    pages = {408-410},
    year = {2009},
    }

@article{Murphy2016,
   author = {Kieran A. Murphy and Nikolaj Reiser and Darius Choksy and Clare E. Singer and Heinrich M. Jaeger},
   issn = {14347636},
   issue = {2},
   journal = {Granul. Matter},
   month = {5},
   pages = {1-9},
   publisher = {Springer New York LLC},
   title = {Freestanding loadbearing structures with Z-shaped particles},
   volume = {18},
   url = {https://link.springer.com/article/10.1007/s10035-015-0600-2},
   year = {2016}
}

@article{Franklin2014,
   author = {Scott V Franklin},
   issn = {0295-5075},
   issue = {5},
   journal = {Europhys. Lett.},
   month = {6},
   pages = {58004},
   publisher = {IOP Publishing},
   title = {Extensional rheology of entangled granular materials},
   volume = {106},
   year = {2014}
}

@article{Trepanier2010,
   author = {M. Trepanier and Scott V. Franklin},
   issn = {15393755},
   issue = {1},
   journal = {Phys. Rev. E},
   month = {7},
   pages = {011308},
   publisher = {American Physical Society},
   title = {Column collapse of granular rods},
   volume = {82},
   year = {2010}
}

@article{Aponte2024,
   author = {David Aponte and Nicolas Estrada and Jonathan Barés and Mathieu Renouf and Emilien Azéma},
   issn = {24700053},
   issue = {4},
   journal = {Phys. Rev. E},
   month = {4},
   pages = {044908},
   pmid = {38755878},
   publisher = {American Physical Society},
   title = {Geometric cohesion in two-dimensional systems composed of star-shaped particles},
   volume = {109},
   year = {2024}
}

@article{1,
    author ="Sarate, Palash S. and Murthy, Tejas G. and Sharma, Prerna",
    title  ="Column to pile transition in quasi-static deposition of granular chains",
    journal  ="Soft Matter",
    year  ="2022",
    volume  ="18",
    issue  ="10",
    pages  ="2054-2059",
    publisher  ="The Royal Society of Chemistry"
    }

@article{Brown2012,
  title = {Strain Stiffening in Random Packings of Entangled Granular Chains},
  author = {Brown, Eric and Nasto, Alice and Athanassiadis, Athanasios G. and Jaeger, Heinrich M.},
  journal = {Phys. Rev. Lett.},
  volume = {108},
  issue = {10},
  pages = {108302},
  numpages = {4},
  year = {2012},
  month = {Mar},
  publisher = {American Physical Society}
}

@article{Dumont2018,
   author = {Denis Dumont and Maurine Houze and Paul Rambach and Thomas Salez and Sylvain Patinet and Pascal Damman},
   issn = {10797114},
   issue = {8},
 pages = {088001},
   journal = {Phys. Rev. Lett.},
   month = {2},
   pmid = {29543026},
   publisher = {American Physical Society},
   title = {Emergent Strain Stiffening in Interlocked Granular Chains},
   volume = {120},
   year = {2018}
}

@book{Wood1991,
   author = {David Muir Wood},
   isbn = {9780521332491},
   publisher = {Cambridge University Press},
   title = {Soil Behaviour and Critical State Soil Mechanics},
   year = {1991}
}

@article{BenNaim2001,
  title = {Knots and Random Walks in Vibrated Granular Chains},
  author = {Ben-Naim, E. and Daya, Z. A. and Vorobieff, P. and Ecke, R. E.},
  journal = {Phys. Rev. Lett.},
  volume = {86},
  issue = {8},
  pages = {1414-1417},
  numpages = {0},
  year = {2001},
  month = {Feb},
  publisher = {American Physical Society}
}

@article{he2025weaving,
  title={Weaving-inspired asymmetric entangled nodes in multi-component polymer networks},
  author={He, Zejian and Chen, Liya and You, Wei and Zhu, Chenkai and Yang, Xue and Mei, Honggang and Xiao, Ding and Song, Qi and Shan, Tianyu and Yu, Wei and others},
  journal={Nat. Mater.},
  pages={107-116},
  year={2026},
volume    = {25},
  number    = {1},
  publisher={Nature Publishing Group UK London}
}

@book{OSullivan2011,
   author = {Catherine O’Sullivan},
   isbn = {9781482266498},
   title = {Particulate Discrete Element Modelling: A Geomechanics Perspective},
   journal = {Particulate Discrete Element Modelling: A Geomechanics Perspective},
   publisher = {CRC Press},
   year = {2011}
}

@article{bolton1986strength,
  title={The strength and dilatancy of sands},
  author={Bolton, MD},
  journal={Geotechnique},
  volume={36},
  number={1},
  pages={65-78},
  year={1986},
  publisher={Thomas Telford Ltd}
}

@article{de1971reptation,
  title={Reptation of a polymer chain in the presence of fixed obstacles},
  author={De Gennes, Pierre Giles},
  journal={The Journal of Chemical Physics},
  volume={55},
  number={2},
  pages={572-579},
  year={1971},
  publisher={American Institute of Physics}
}

@article{doi1978dynamics,
  title={Dynamics of concentrated polymer systems. Part 1. Brownian motion in the equilibrium state},
  author={Doi, Masao and Edwards, S. F.},
  journal  ="J. Chem. Soc.{,} Faraday Trans. 2",
  volume={74},
  pages={1789-1801},
  year={1978},
  publisher={The Royal Society of Chemistry}
}

@article{mackintosh1995elasticity,
  title={Elasticity of semiflexible biopolymer networks},
  author={MacKintosh, FC and K{\"a}s, Josef and Janmey, PA},
  journal={Phys. Rev. Lett.},
  volume={75},
  number={24},
  pages={4425},
  year={1995},
  publisher={APS}
}

@book{larson1999structure,
  title={The Structure and Rheology of Complex Fluids},
  author={Larson, R.G.},
  isbn={9780195121971},
  lccn={98019940},
  series={Topics in Chemical Engineering},
  year={1999},
  publisher={OUP USA}
}

@article{pivlab,
    title = "PIVlab – Towards User-friendly, Affordable and Accurate Digital Particle Image Velocimetry in MATLAB",
    author = "Eize Stamhuis and William Thielicke",
    year = "2014",
    month = oct,
    day = "16",
    pages = {e30},
    volume = "2",
    journal = "Journal of Open Research Software",
    number = "1",
}

@article{bhat2023force,
  title={Force transmission during repose of flexible granular chains},
  author={Bhat, Mohd Ilyas and Sharma, Prerna and Sitharam, TG and Murthy, Tejas G},
  journal={Soft Matter},
  volume={19},
  number={44},
  pages={8493--8506},
  year={2023},
  publisher={Royal Society of Chemistry}
}

@article{pena2007influence,
  title={Influence of particle shape on sheared dense granular media},
  author={Pena, AA and Garcia Rojo, R and Herrmann, Hans J{\"u}rgen},
  journal={Granul. Matter},
  volume={9},
  number={3},
  pages={279--291},
  year={2007},
  publisher={Springer}
}

@article{
Mandal2020,
author = {Sandip Mandal  and Maxime Nicolas  and Olivier Pouliquen },
title = {Insights into the rheology of cohesive granular media},
journal = {Proc. Natl. Acad. Sci.},
volume = {117},
number = {15},
pages = {8366-8373},
year = {2020},
doi = {10.1073/pnas.1921778117},
}

@article{
Jung2025,
author = {Yeonsu Jung  and Thomas Plumb-Reyes  and Hao-Yu Greg Lin  and L. Mahadevan },
title = {Entanglement transition in random rod packings},
journal = {Proc. Natl. Acad. Sci.},
volume = {122},
number = {8},
pages = {e2401868122},
year = {2025},
doi = {10.1073/pnas.2401868122},
}

@article{
Karuriya2023,
author = {Ashta Navdeep Karuriya  and Francois Barthelat },
title = {Granular crystals as strong and fully dense architectured materials},
journal = {Proc. Natl. Acad. Sci.},
volume = {120},
number = {1},
pages = {e2215508120},
year = {2023},
doi = {10.1073/pnas.2215508120},
}

@article{
Haver2024,
author = {Daan Haver  and Daniel Acuña  and Shahram Janbaz  and Edan Lerner  and Gustavo Düring  and Corentin Coulais },
title = {Elasticity and rheology of auxetic granular metamaterials},
journal = {Proc. Natl. Acad. Sci.},
volume = {121},
number = {14},
pages = {e2317915121},
year = {2024},
doi = {10.1073/pnas.2317915121},
}

@article{Cao2018,
  author = {Cao, Yixin and Li, Jindong and Kou, Binquan and Xia, Chengjie and Li, Zhifeng and Chen, Rongchang and Xie, Honglan and Xiao, Tiqiao and Kob, Walter and Hong, Liang and Zhang, Jie and Wang, Yujie},
  title = {Structural and topological nature of plasticity in sheared granular materials},
  journal = {Nat. Commun.},
  year = {2018},
  volume = {9},
  number = {1},
  pages = {2911},
  doi = {10.1038/s41467-018-05329-8}
}

@article{
Juanes2024,
author = {Wei Li  and Ruben Juanes },
title = {Dynamic imaging of force chains in 3D granular media},
journal = {Proc. Natl. Acad. Sci.},
volume = {121},
number = {14},
pages = {e2319160121},
year = {2024},
doi = {10.1073/pnas.2319160121},
}

\end{document}